\documentclass[%
 aip,
 amsmath,amssymb,
 reprint,%
]{revtex4-1}

\usepackage{graphicx}
\usepackage{dcolumn}
\usepackage{bm}

\usepackage[utf8]{inputenc}
\usepackage[T1]{fontenc}
\usepackage{mathptmx}

\begin{document}

\preprint{AIP/123-QED}

\title{Assembly of particle strings via isotropic potentials}

\author{D. Banerjee}
\affiliation{McKetta Department of Chemical Engineering, University of Texas at Austin, Austin, Texas 78712, USA}
\author{B. A. Lindquist}
\affiliation{McKetta Department of Chemical Engineering, University of Texas at Austin, Austin, Texas 78712, USA}
\author{R. B. Jadrich}
\affiliation{McKetta Department of Chemical Engineering, University of Texas at Austin, Austin, Texas 78712, USA}
\author{T. M. Truskett} 
 \email{truskett@che.utexas.edu}
 \affiliation{McKetta Department of Chemical Engineering, University of Texas at Austin, Austin, Texas 78712, USA}
 \affiliation{Department of Physics, University of Texas at Austin, Austin, Texas 78712, USA}

\date{\today}

\begin{abstract}

Assembly of spherical colloidal particles into extended structures, including linear strings, in the absence of directional interparticle bonding interactions or external perturbation could facilitate the design of new functional materials. Here, we use methods of inverse design to discover isotropic pair potentials that promote the formation of single-stranded, polydisperse strings of colloids "colloidomers" as well as size-specific, compact colloidal clusters. Based on the designed potentials, a simple model pair interaction with a short-range attraction and a longer-range repulsion is proposed which stabilizes a variety of different particle morphologies including (i) dispersed fluid of monomers, (ii) ergodic short particle chains as well as porous networks of percolated strings, (iii) compact clusters, and (iv) thick cylindrical structures including trihelical Bernal spirals.

\end{abstract}

\maketitle

\section{Introduction} 
Owing to the chemical and structural versatility of their building blocks, colloidal materials can be designed to assemble into a variety of microstructures. One motif that may enable functionality is chains or strings of colloids (so-called "colloidomers"\cite{DropletFJC}). When percolated, these structures can serve as a template for conductivity, facilitating photonic and electronic transfer along the connected interparticle network.~\cite{Tang1D, Velev} Mechanically stable particle networks with high surface-to-volume ratio are also interesting morphologies for nanoporous catalysts,~\cite{Catalyst, catalysis} and open network structures may be advantageous for applications that require materials that can dynamically reconfigure in response to a stimulus.

Self-assembling colloidal strings have typically been designed by choosing building blocks with anisotropic interactions commensurate with the targeted morphology. For instance, thin chains of particles have been assembled \emph{in silico} from hard spheres, each decorated with two colinear attractive patches.~\cite{PatchyDuguet, SciortinoJCP2007, BethLinker, SciortinoPatchy} Such directionally specific interactions have been engineered in practice by grafting appropriate functional groups (e.g., DNA) to the surface of colloids.~\cite{DropletFJC, Jasna, PBS} Analogous physics can be achieved via short-range anisotropic dipolar interactions, which have been shown to promote linear chain growth between charged gold nanoparticles.~\cite{chargeAu} Strong dipole interactions have been argued to be the driving force behind the self-organization of other nanoparticles into one-dimensional chains~\cite{Tang237, Dipole1, Dipole2} and three-dimensional percolated fractal chain networks.~\cite{Lin3D} 

Anisotropic colloid shape can similarly be tailored to obtain flexible colloidal chains or stringy structured fluids.~\cite{LockKey,Polloids} For instance, Sacanna and coworkers employed Fischer's lock and key recognition mechanism between a homogeneous sphere and a sphere with a cavity to assemble compact clusters as well as more complex and flexible colloidal polymers. Particle and interaction anisotropy can also be induced by the assembly process itself. Under certain experimental conditions, spherical and uniformly grafted nanoparticles in a homopolymer matrix self-assembled into linear chains.~\cite{Akcora} The short-ranged depletion attraction in these systems is argued to be counterbalanced by the entropy of distortion which arises when the grafted brushes on two nanoparticles compress due to steric constraints upon approach. This can lead to an anisotropic distribution of the local graft density~\cite{Akcora, Arya} imparting an amphiphilic character to the nanoparticles. Such anisotropic assembly of uniformly grafted nanoparticles has also been predicted via theory as well as simulations where the ligands are modeled explicitly.~\cite{Bedrov, Samanvaya, Jiao, Koerner, Victor, Lafitte, Pana}

A few attempts have been made to design an isotropic pair potential that causes a single-component fluid of particles to self-assemble into stringlike structures. To this end, Rechtsman et al~\cite{Rechtsman} proposed a complex "five-finger potential" containing five repeating attractive wells at intervals set by the particle diameter that are separated by repulsive barriers that inhibit formation of compact objects. 
In two dimensions, simpler potentials, though still possessing competing attractions and repulsions, have been shown to generate stringy structures. For example, coarse-graining multicomponent simulations of grafted nanoparticles revealed several single-component isotropic pairwise potentials that promote self-assembly of distinct morphologies: dispersed particles, long strings, and a percolated network.~\cite{Lafitte} Moreover, a class of potentials characterized by a single attractive well followed by a repulsive barrier furnished by a piecewise function of linear components has displayed different microstructures, ranging from monomers to aggregates to short strings to a labyrinthine chain network, as a function of area fraction and range of the repulsion.~\cite{haw} 

Common to the above studies is the presence of competing interactions, such as a short-range attractive and long-range repulsive (abbreviated SALR) potential, also known to promote the self-assembly of more compact particle clusters. The repulsive interactions in such potentials naturally limit the aggregation that would otherwise be promoted by the attractions. Self-assembly of clusters and strings requires growth that is self-limited--for clusters with respect to their overall size and for strings with respect to the dimension normal to growth. A few studies further reinforce potential connections between compact and elongated cluster morphologies. For instance, it was shown in both experiment~\cite{Bartlett} and simulation~\cite{Sciortino1D,Bolhuis} that both clusters and thick ramified structures are possible when systems possess competing SALR interactions. Similarly, in simulation, an SALR potential was shown to produce percolating states with a mixture of transient filamentous and spherical aggregates when the packing fraction exceeded $0.148$.~\cite{Cardinaux} 

Motivated by the investigations described above, the aim of this paper is two-fold. The first is to use methods of inverse design~\cite{Torquato, Inverse-Jain} (specifically a recent strategy~\cite{RyanRE} based on relative entropy coarse-graining~\cite{Shell,Shell2}) to discover isotropic potentials that promote self-assembly of one-monomer wide chains or compact clusters, respectively, in a one-component system of spherical particles. The second aim is to identify, on the basis of the designed interactions, a simpler model pair potential that favors assembly of these and related structures as a function of the length scales of the competing attractive and repulsive interactions.  

The balance of this paper is structured as follows. Sect.~\ref{sec:methods} outlines the relative entropy based method we adopt for inverse design and presents details of the molecular simulations. The pair potentials and structures resulting from the inverse design  for both compact clusters and strings are described in Sect.~\ref{sec:IO}. Motivated by the qualitative forms of the designed interactions, Sect.~\ref{sec:UP} introduces a simpler related pair potential and explores the various morphologies that it favors as a function of its parameters.
Conclusions and possible directions for future research are presented in Sect.~\ref{sec:conclusions}.

\section{Methods}
\label{sec:methods}

    \subsection{Relative Entropy Coarse-Graining} 
    \label{subsec:RE}
    
    Relative entropy (RE) coarse-graining,~\cite{Shell,Shell2} also known as likelihood maximization in probability and statistics, is used in this work to obtain isotropic potentials capable of self-assembling particles into different target structures. Commonly applied to obtain a reduced dimensionality description of complex molecules for simulation, RE course-graining has more recently been used to design isotropic pair interactions that lead to self-organization of a rich variety of equilibrium structures including fluidic clusters,~\cite{RyanIC,RyanRE} porous mesophases,~\cite{BethPores,RyanRE} and crystalline lattices.~\cite{BL_RE,BethFK,RyanRE} 
    
    In brief, the RE course-graining protocol considers a target ensemble of particle configurations that collectively exhibits a desired structural motif (e.g., strings or compact clusters), discussed below.
    The optimized interactions are those that maximize the overlap of the probability distribution for configurations at equilibrium with that of the target ensemble. Here, we consider an isotropic pair potential, $U(r|\boldsymbol{\theta})$, with $m$ tunable parameters $\boldsymbol{\theta} = [\theta_{1}, \theta_{2}, \cdots, \theta_{m}]$. According to RE coarse-graining, the parameters are updated in an iterative manner via   
    \begin{equation} 
        \label{eqn:RE_eqn}
        \boldsymbol{\theta}^{k+1} = \boldsymbol{\theta}^{k} + \alpha \int_{0}^{\infty} r^2 [g(r|\boldsymbol{\theta}^{k}) - g_{\text{tgt}}(r)][\nabla_{\boldsymbol{\theta}}{\beta}U(r|\boldsymbol{\theta})]_{\boldsymbol{\theta} = \boldsymbol{\theta}^{k}} dr
    \end{equation} 
    where $\beta=(k_{B}T)^{-1}$, $k_{B}$ is the Boltzmann constant, $T$ is temperature, $\alpha$ is the learning rate, $g(r|\boldsymbol{\theta}^{k})$ is the radial distribution function of the system in the $k^{th}$ iterative step of the optimization, and $g_{\text{tgt}}(r)$ is the radial distribution function of the target ensemble. In practice, $g(r|\boldsymbol{\theta}^{k})$ is obtained from the equilibrium particle configurations of a molecular simulation using $U(r|\boldsymbol{\theta}^{k})$.~\cite{RyanRE} A rigorous mathematical derivation of the above update scheme is reviewed in Refs.~\citenum{RyanRE},~\citenum{BL_RE}, and~\citenum{William}. The outcome of a successful optimization is a thermally non-dimensionalized interaction ${\beta}U_{{\text opt}}(r)$ that results in an equilibrium structure that closely mimics that of the target ensemble.
    
    \subsection{Design of Targets} 
    \label{subsec:target}
    The first step in the inverse design protocol described above is the construction of an ensemble of target configurations from which $g_{\text{tgt}}(r)$ can be computed. Target distributions can be specified in any way that yields an ensemble of desired configurations with convergent statistics. Here, we employ equilibrium statistical mechanics via canonical-ensemble molecular dynamics simulations with $N$ spherical particles of diameter $\sigma$ in a periodically replicated cubic cell of side length $L$ [i.e., packing fraction $\eta = N \pi \sigma^3/(6L^3)$] at temperature $T$. Dimensionless (generally many-body) interparticle potentials $\beta V$, given below, are selected for each target ensemble to yield configurations characteristic of the desired morphology.

        \subsubsection{Clusters} 
        \label{subsec:tgt_cluster}
        Target ensembles of monodisperse, compact clusters of size $N_{\text{tgt}} = 2$ (dimer), $4$ (tetramer) and $8$ (octamer) are generated via molecular dynamics simulations in the canonical ensemble at packing fraction $\eta = 0.025$. The following interactions are chosen to mimic the desired target morphology. Particles of diameter $\sigma$ interact with hard-sphere-like repulsions represented via a Weeks-Chandler-Andersen (WCA) potential $V_{\text{\tiny{WCA}}}(r)$. 
        \begin{equation}
        \label{eqn:WCA}
        V_{\text{\tiny{WCA}}}(r) = \left\{
                        \begin{array}{ll}
                            4 {\varepsilon_{w}}\bigg[ \bigg(\dfrac{\sigma}{r} \bigg)^{12} - \bigg( \dfrac{\sigma}{r} \bigg)^{6} \bigg] + {\varepsilon_{w}}, & r \leq 2^{1/6}\sigma \\ [15pt]
                            
                            0, &  r > 2^{1/6}\sigma \\
                        \end{array} 
                    \right.
        \end{equation}
        where $\beta\varepsilon_{w} = 5$. Particles are assigned to a particular cluster, the compactness of which is enforced by applying an additional finitely extensible non-linear elastic spring (FENE) potential $V_{\text{\tiny{FENE}}}$ between each particle in the cluster
        \begin{equation}
        \label{eqn:FENE}
        V_{\text{\tiny{FENE}}}(r) = \left\{
                        \begin{array}{ll}
                            -\dfrac{1}{2}kr_{0}^{2}\text{ln}\bigg[1 - \bigg( \dfrac{r}{r_{0}} \bigg)^{2} \bigg], & r \leq r_{0} \\ [15pt]
                            
                            \infty, &  r > r_{0} \\
                        \end{array} 
                    \right.
        \end{equation}
        with $k$ = $30 k_{\text{B}}T/\sigma^{2}$, and $\text{r}_0 = 1.5\sigma$ ($N_{\text{tgt}} = 2$) or $5.0\sigma$ ($N_{\text{tgt}} = 4, 8$). Additionally, a minimum distance of separation between the clusters is ensured by introducing an isotropic Yukawa repulsion between particles in different clusters 
        \begin{equation}
        \label{eqn:Yukawa}
        V_{\text{\tiny{Yukawa}}}(r) = \left\{
                        \begin{array}{ll}
                            \varepsilon_{y}\dfrac{\text{exp}(-\kappa{r})}{r}, & r < r_{cut} \\ [15pt]
                            
                            0, &  r \geq r_{cut} \\
                        \end{array} 
                    \right.
        \end{equation}
        where $\beta\varepsilon_{y} =$ 30, 10, and 0 for $N_{\text{tgt}} =$ 2, 4, 8, respectively, $\kappa = 0.5{\sigma}^{-1}$, and $r_{cut} = L/2.5$. After ensuring equilibration, the target radial distribution function $g_{\text{tgt}}(r)$ for $N_{\text{tgt}} = 2$, $4$, $8$ sized cluster fluids is computed.

        \subsubsection{Strings} 
        \label{subsec:tgt_string}
        For strings, we created the target ensemble from simulations of linear particle chains of molecular weight $N_{\text{tgt}}$ = 10. Monomers interact via the repulsive WCA potential of Eq.~\ref{eqn:WCA} with $\beta\varepsilon_{w} = 1$, and adjacent beads interact via the FENE spring potential $V_{\text{\tiny{FENE}}}$ of Eq.~\ref{eqn:FENE} with $k = 30 k_{\text{B}}T/\sigma^{2}$ and $r_{0} =  1.5\sigma$. Non-bonded monomers also interact with the Yukawa potential $V_{\text{\tiny{Yukawa}}}$ of Eq.~\ref{eqn:Yukawa} with $\beta\varepsilon_{y} = 30$, $\kappa = 0.5{\sigma}^{-1}$, and $r_{cut} = L/3.0$. The system of linear chains is allowed to evolve via molecular dynamics and, upon equilibration, $g_{\text{tgt}}(r)$ is calculated at different packing fractions $\eta$ = 0.05, 0.1 and 0.15.

    \subsection{Simulation Details} 
    \label{simulation}

    Molecular dynamics simulations for the target structures and the optimization are performed in the canonical ensemble with a periodically replicated cubic simulation cell. 
    
    The software package HOOMD-blue 2.3.4~\cite{Hoomd1, Hoomd2} is used to generate the target configurations of particles of mass $m$. A time step of $dt = 0.005\sqrt{\sigma^{2}m/k_{B}T}$ is adopted and the Nos\`{e}-Hoover thermostat with a time constant of $\tau = 50dt$ is employed.
    For target compact clusters of $N_{\text{tgt}} = 2$, $4$, and $8$ at a packing fraction of $\eta = 0.025$, $N = 384$ particles are used in a periodic cell of size $L = 20\sigma$.
    Target structures for strings are generated using $N = 320$, $420$ and $630$ particles in cubic cells of dimension $L = 15\sigma$, $13\sigma$, and $13\sigma$ for different packing fractions of $\eta$ = 0.05, 0.10 and 0.15. 
    
    Simulations within the iterative RE optimization as well as for the forward runs with the optimized potentials were performed in GROMACS 5.1.2 with a time step of $dt = 0.001\sqrt{\sigma^{2}m/k_{B}T}$. Constant temperature was maintained using the velocity-rescale thermostat with a time constant $\tau = 50dt$. The phase diagram of Sect.~\ref{sec:UP} was generated using $N = 968$ particles in a box size $L = 15\sigma$ (i.e., $\eta = 0.15$), collecting configurations for $3\times10^7$ time steps. In order to extract improved cluster statistics as well as to check for finite-size effects, simulations with the larger box size of $L = 30\sigma$ with 7744 particles were also performed for select state points. For these same state points, simulations were also conducted with different initial configurations (randomly placed particles, compact aggregates, and linear rod-like clusters) to test the robustness and sensitivity of the final structures as well as to ensure equilibrium is attained. The simulation snapshots were rendered using the Visual Molecular Dynamics~\cite{vmd} software. 
    
    \subsection{Structural Analysis} 
    \label{subsec:StructuralAnalysis}
    The structures obtained from the simulations described above are characterized by the cluster size distribution (CSD), the distribution of the number of nearest neighbors, the fractal dimension of the resulting aggregates $d_f$, and a percolation analysis. The CSD quantifies the fraction of clusters that contain $n$ particles, where a particle is considered to belong to a cluster if its center is within a prescribed cut-off distance $r_{\text{cut}}$ from the center of at least one other particle in the same cluster. The smallest range of attraction for the optimized potentials in this work is $\sim$ $1.1\sigma$ which makes it a natural choice for $r_{\text{cut}}$. For both the case of spherical clusters and strings, various cut-offs [$1.05\sigma - 1.25\sigma$] yield non-perceptible changes in the CSD and other structural properties. 
    
    To distinguish between thick ramified structures and "thin" chains of one-monomer diameter width, the number of nearest neighbors of each particle is evaluated. Single strands of strings have predominantly two bonds per particle. Additionally, to characterize the anisotropy of the resulting objects, the fractal dimension $d_{f}$ is determined via $R_{g} \sim n^{1/d_{f}}$,
    where $n$ is the number of particles in a cluster and $R_g$ is the radius of gyration. The latter for a cluster of size $n$ is defined as 
    \begin{equation} \label{eqn:Rg}
    R_g(n) = \dfrac{1}{n^{1/2}} {\Bigg \langle {\Big[ \sum\limits_{i=1}^n {(\textbf{r}_{i} - \textbf{R}_{\text{CM}})^{2}} \Big]}} \Bigg \rangle^{1/2}
    \end{equation} 
    where $\textbf{r}_{i}$ and $\textbf{R}_{\text{CM}}$ are the coordinates of the $i^{\text{th}}$ particle and the center of mass of the cluster of $n$ particles, respectively. The average is performed over all clusters of size $n$. The fractal dimension is expected to be bounded by two limits: $\sim$3 for compact, homogeneously spherical aggregates to $\sim$1 for linear objects. For stringlike objects, we sometimes find that a single value for $d_f$ does not satisfactorily fit the data for every aggregate size $n$, in which case we segregate the data that visually appear to have different slopes in the $R_g$ versus $\text{log}(n)$ plot and compute distinct values for $d_f$ for the different regions of $n$.
    
    Some of the structures discussed below are found to be percolating. Percolation is defined if a cluster spans the length of the box in at least one direction such that, under periodic boundary conditions, the cluster wraps around the box and connects to itself. Accounting for periodicity, a percolating cluster is thus infinitely long and hence the corresponding $d_f$ or $R_g$ is not computed for percolated structures. If at least 50\% of the configurations conform to the said definition, the resulting morphology is deemed to be percolating.~\cite{BethLinker}

\section{Inverse Design}
\label{sec:IO}
    \subsection{Compact clusters} 
    \label{subsec:IO_Clusters}
    Given the connection between clusters and stringlike structures described in the Introduction, it is instructive to use inverse design to discover pair potentials that favor these morphologies. As described in Sect.~\ref{sec:methods}, we first use RE optimization here to determine isotropic pair potentials that lead to self-assembly of compact $N_{\text{tgt}}$-mer clusters, namely, dimers, tetramers and octamers (target sizes of $N_{\text{tgt}}$ = 2,4,8) at a packing fraction of $\eta$ = 0.025, where the choice of $\eta$ is motivated by prior work.~\cite{RyanIC} In each case, the RE optimization discovers an interaction ${\beta}U_{{\text opt}}(r)$ capable of successfully assembling the target morphology, despite not quantitatively reproducing $g_{\text{tgt}}(r)$.  For example, the region of depletion in $g(r)$ due to the repulsion between separate clusters is less pronounced in the equilibrium assembled structure relative to the target ensemble, and the distinct crystalline peaks for the target octamers are mimicked by a more muted profile in the equilibrium assembly~\cite{RyanIC}; a detailed comparison is provided in the Appendix (Fig. A1). 
    
    Fig.~\ref{fgr:RE_cluster}a shows the optimized pair potentials ${\beta}U_{\text{opt}}(r)$ that form such compact clusters. The two main features of the optimized potentials are the attractive well beginning at $r=\sigma$ followed by a repulsive barrier. In accord with prior work,~\cite{RyanIC} the magnitude of the attraction increases and the peak of the repulsive barrier shifts to larger separations with increasing aggregate size. Using the designed pairwise interaction, strong clustering emerges. The corresponding CSDs in Fig.~\ref{fgr:RE_cluster}b demonstrate that the target cluster sizes are reproduced for all three cases of $N_{\text{tgt}}$ = 2, 4 and 8. There is mild polydispersity with respect to aggregation number, though most clusters are within one number of the targeted size. The self-assembled aggregates are well-separated and behave as an equilibrium cluster fluid. See a snapshot for $N_{\text{tgt}} = 8$ in Fig.~\ref{fgr:RE_cluster}c. The compactness of the largest clusters ($N_{\text{tgt}} = 8$) is quantified by the corresponding fractal dimension $d_{f}$, which is given by the inverse of the slope of Fig.~\ref{fgr:RE_cluster}(d) and is approximately 2.9. 
    
    \begin{figure}
        \includegraphics[width=3.37in,keepaspectratio]{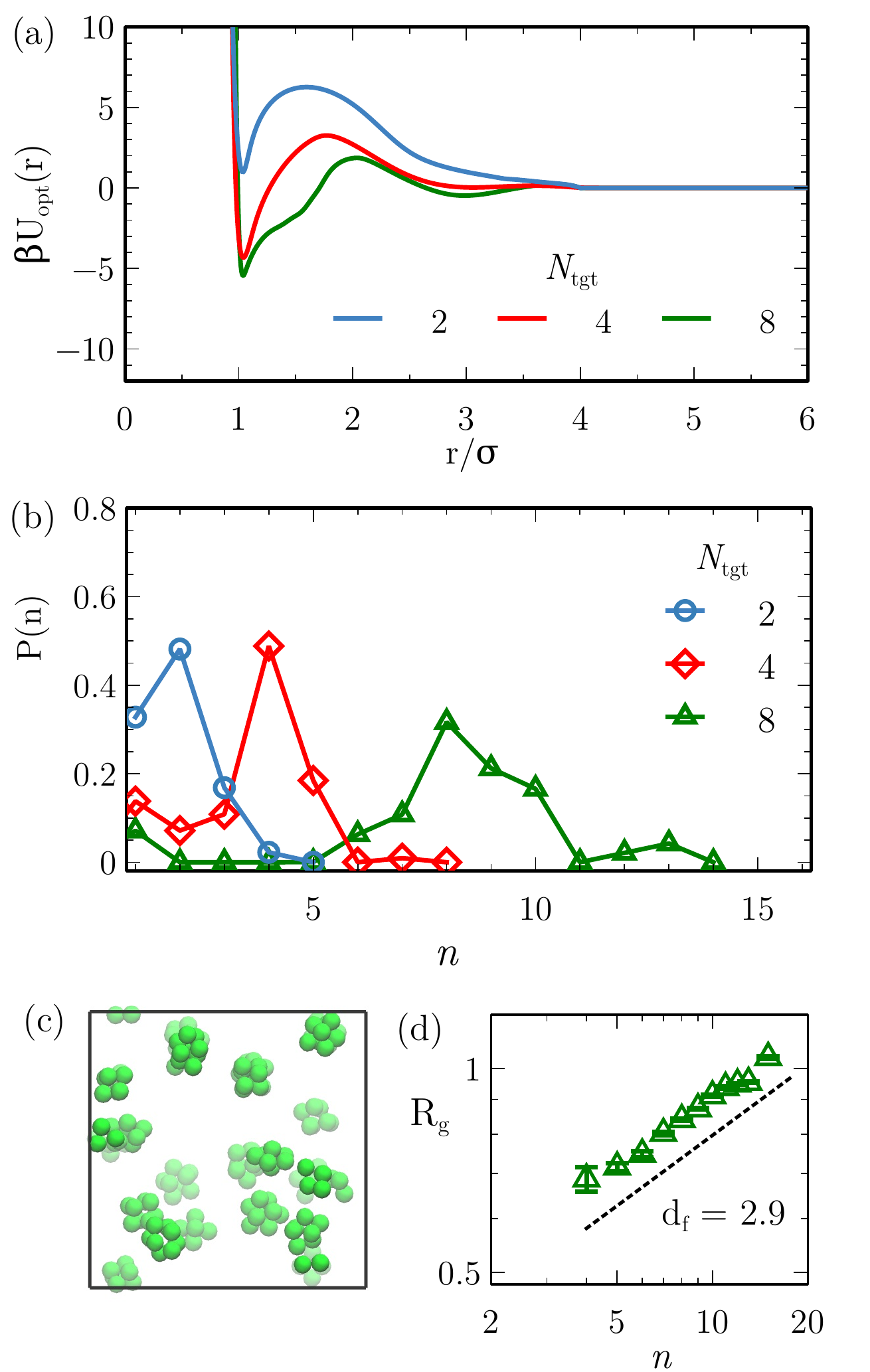}
        \caption{(a) Optimized potentials ${\beta}U_{\text{opt}}(r)$ obtained through inverse design of compact $N_{\text{tgt}}$-mer clusters of particles at packing fraction $\eta = 0.025$. (b) Fraction of clusters $P(n)$ containing $n$ particles, using the above optimized potentials. (c) Simulation snapshot at equilibrium for the potential designed to assemble $N_{\text{tgt}} = 8$ compact clusters. (d) The average radius of gyration $R_g$ of spherical clusters of $N_{\text{tgt}} = 8$, with their corresponding error bars, as a function of cluster size $n$ on a log-log plot. The fractal dimension is the inverse of the slope ($R_{g} \sim n^{1/d_{f}}$) and is found to be approximately 2.9.}
        \label{fgr:RE_cluster}
        \end{figure}
    
    The $N_{\text{tgt}} = 2$ case of self-assembling dimers shares some complexities with the problem of string formation in that growth must be limited to a single direction. A pair of particles must associate attractively at near $r=\sigma$, but formation of triangles, where an incoming particle bonds between the dimer pair at the point of closest approach to both centers (i.e., $r=\sigma$), must be suppressed. Indeed, a dimer can be considered as both the smallest cluster and the smallest string. Looking forward to self-assembly of strings, we might anticipate that the characteristics of the potential optimized for forming dimers (a net-repulsive potential with a relatively narrow attractive well) are favorable for string formation more generally.  
    
    \subsection{Strings} 
    \label{subsec:IO_Strings}
    
    As described in Sect.~\ref{sec:methods}, the target ensemble of configurations for particle strings considered here comprises chains of particles of molecular weight $N_{\text{tgt}}=10$. In general, the isotropic pair potentials resulting from the corresponding RE optimizations ${\beta}U_{{\text opt}}(r)$ successfully self-assemble fluids of stringy particles. The radial distribution functions $g(r)$ of particles interacting via the optimized potentials capture the salient features of the target structure $g_{\text{tgt}}(r)$, though--as shown in the Appendix (Fig. A2)--the depleted region present between $r=\sigma$ and $r=2\sigma$ due to the absence of compact aggregates is less prominent in the former compared to the latter.
    
\begin{figure}
        \includegraphics[width=3.37in,keepaspectratio]{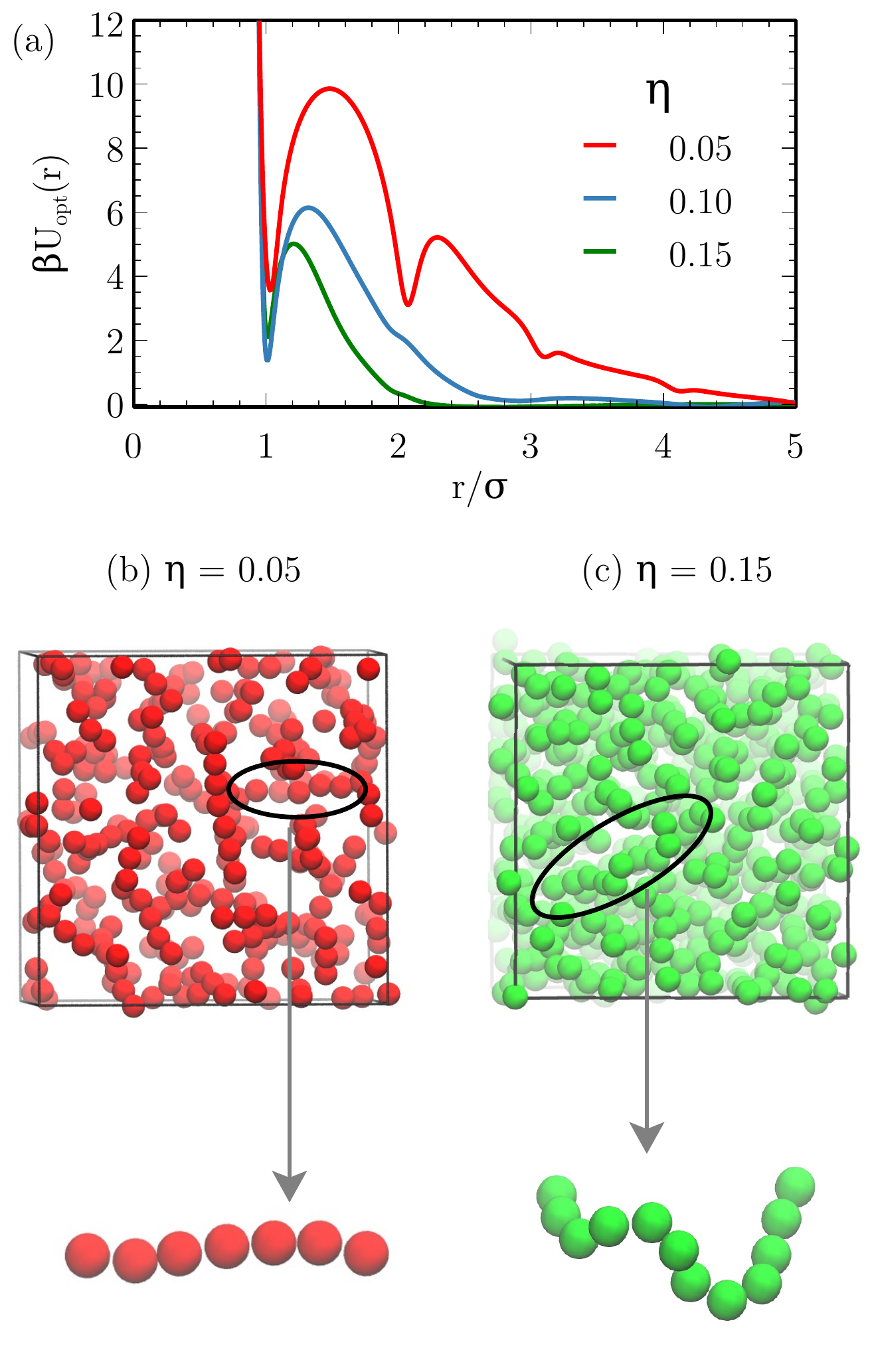}
        \caption{(a) Optimized isotropic potentials ${\beta}U_{{\text opt}}(r)$ obtained by the inverse design of linear chains at three different packing fractions $\eta$. (b \& c) Simulation snapshot at equilibrium of strings formed using ${\beta}U_{{\text opt}}(r)$ at $\eta$ = 0.05 and 0.15, respectively.}
        \label{fgr:RE_string_potn}
    \end{figure}

    Fig.~\ref{fgr:RE_string_potn} shows ${\beta}U_{{\text opt}}(r)$ at three different packing fractions $\eta$ = 0.05, 0.1 and 0.15. Analogous to the potential optimized for forming dimers in Sect.~\ref{subsec:IO_Clusters}, the attractive well is relatively narrow for all $\eta$. The repulsive barrier, on the other hand, is more sensitive to the packing fraction, increasing in range and magnitude as $\eta$ is reduced. For $\eta$ = 0.05, the pair potential is more complex due to the emergence of secondary features on the scale of a monomer diameter $\sigma$. The repulsive hump for $\eta$ = 0.05 is followed by three secondary attractive minima at $r$ = $2\sigma$, $3\sigma$ and $4\sigma$ respectively. This potential with alternating attractions and repulsions is qualitatively reminiscent of the "five-finger potential" proposed in  Ref.~\citenum{Rechtsman} to form colloidal strings. These features are progressively muted as the packing fraction of the target structure and the optimization is increased. For $\eta$ = 0.1, there is a slight hint of a dip at $r = 2\sigma$ which is further reduced for $\eta$ = 0.15 where the repulsion terminates at $r = 2.25\sigma$.
    
\begin{figure}
        \includegraphics[width=3.37in,keepaspectratio]{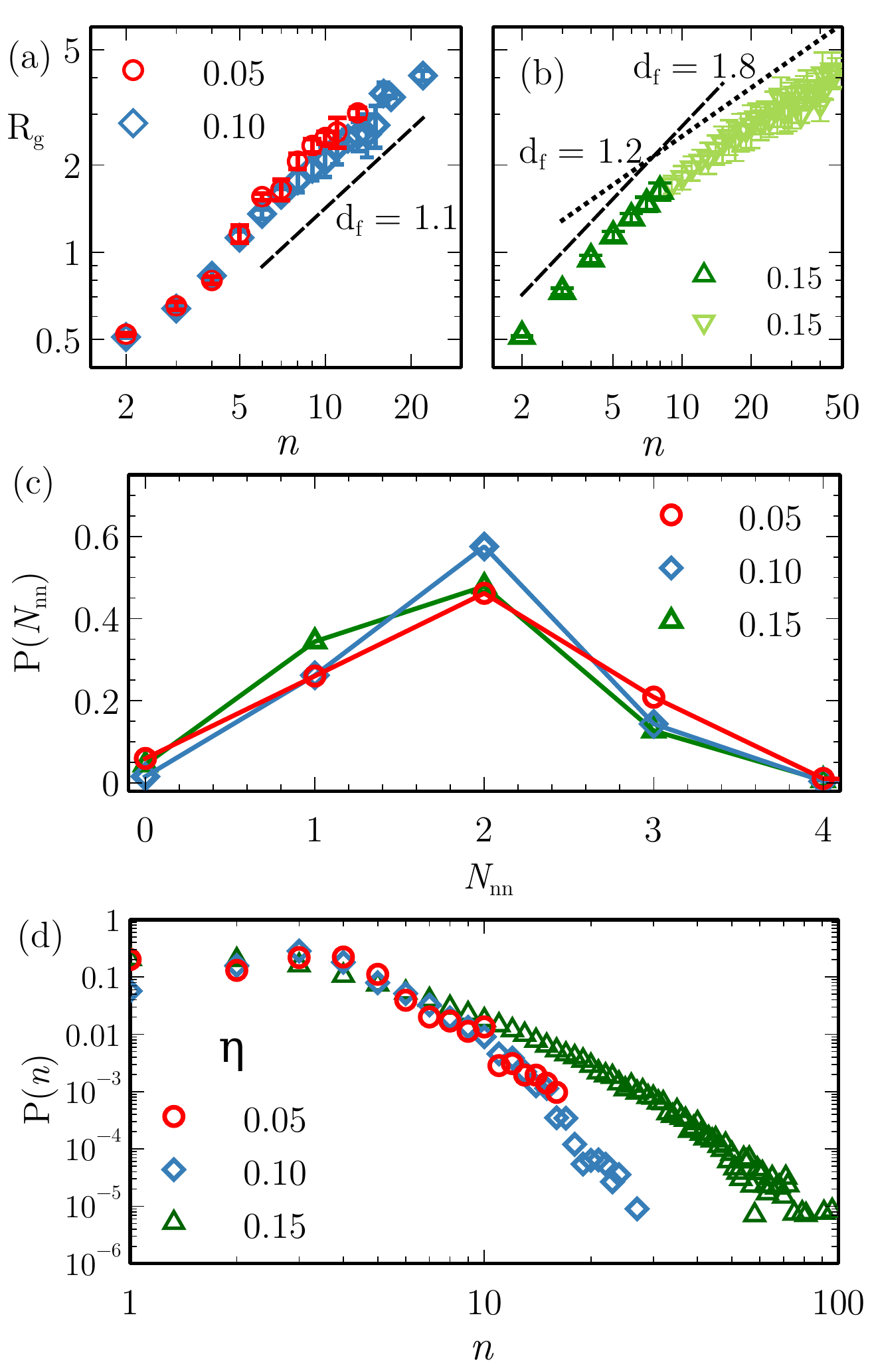}
        \caption{(a and b) Average radius of gyration of the clusters self-assembled from the pair potentials of Fig.~\ref{fgr:RE_string_potn} as a function of cluster size. Fractal dimension $d_f$, computed using $R_{g} \sim n^{1/d_{f}}$, is approximately 1.10 for $\eta$ = 0.05 and 0.10. For $\eta$ = 0.15, $d_f \sim$ 1.20 for strings of length $n \leq 8$ and 1.80 for $n > 8$. (c) Nearest neighbour distribution and (d) cluster size distribution of the aggregates obtained through the use of the optimized isotropic potentials of Fig.~\ref{fgr:RE_string_potn} at the specified volume fractions.}
        \label{fgr:RE_string_CSD}
    \end{figure}

    Using the optimized pair potentials, linear stringlike structures are observed to form at all three packing fractions. Fig.~\ref{fgr:RE_string_potn}(b) and (c) show representative simulation snapshots in equilibrium for $\eta$ = 0.05 and 0.15, where stringlike objects are visually apparent. More quantitatively, $R_g$ as a function of string size at three different packing fractions are reported in Figs.~\ref{fgr:RE_string_CSD}a,b. The corresponding fractal dimension for the chainlike structures at $\eta$ = 0.05 and 0.10 is approximately 1.1. The fractal analysis at $\eta$ = 0.15 shows two power laws, $d_f \sim 1.20$ for $n \leq 8$ and $1.80$ for $n > 8$, implying that the shorter chains are more linear while the longer strings are more curvilinear. Thus, the resulting optimized aggregates span from rod-like to chain-like, with the linearity of the chains increasing as $\eta$ decreases. Figs.~\ref{fgr:RE_string_potn}b,c highlight two illustrative examples where it is apparent that the selected chain at $\eta = 0.05$ is more linear than that at $\eta = 0.15$.
    
    The chains, at all three volume fractions, predominantly have two nearest neighbours $P(N_{nn})\approx2$; see Fig.~\ref{fgr:RE_string_CSD}b. It may be somewhat surprising that there is such a large percentage of particles ($\approx 20\%$) with three neighbors, particularly for $\eta=0.05\text{,}$ where the $d_f$ indicates that the objects are nearly linear. By considering the size $n$ and $N_{nn}$ for every aggregate, we discern that a fraction of aggregates are actually compact clusters. For example, compact tetramers are characterized by $N_{nn}=3$. Indeed, we find that when the compact clusters are removed from calculation of $P(N_{nn})$, the value of $P(N_{nn}=3)$ is significantly reduced; see Fig. A3 in the Appendix. The percentage of aggregates that are compact clusters at $\eta$ = 0.05, 0.10 and 0.15 are 28.1\%, 20.1\% and 6.7\%, respectively.
    
    Finally, unlike the previous case of compact clusters where distinct peaks are formed at the desired target cluster size from the optimized interactions, here we note no such size-specific assembly for the strings. The CSDs for the optimized potentials are shown in Fig.~\ref{fgr:RE_string_CSD}c, where polydispersity is obviously high and increases with packing fraction. Thus, while our optimization procedure illustrates that isotropic pair potentials can readily assemble monomer-wide stringlike particle structures, they are limited in their ability to control their length.

    \subsection{Comparison of Optimized Interactions} 
    \label{subsec:IO_HeatMap}
   
\begin{figure}
        \includegraphics[width=3.37in,keepaspectratio]{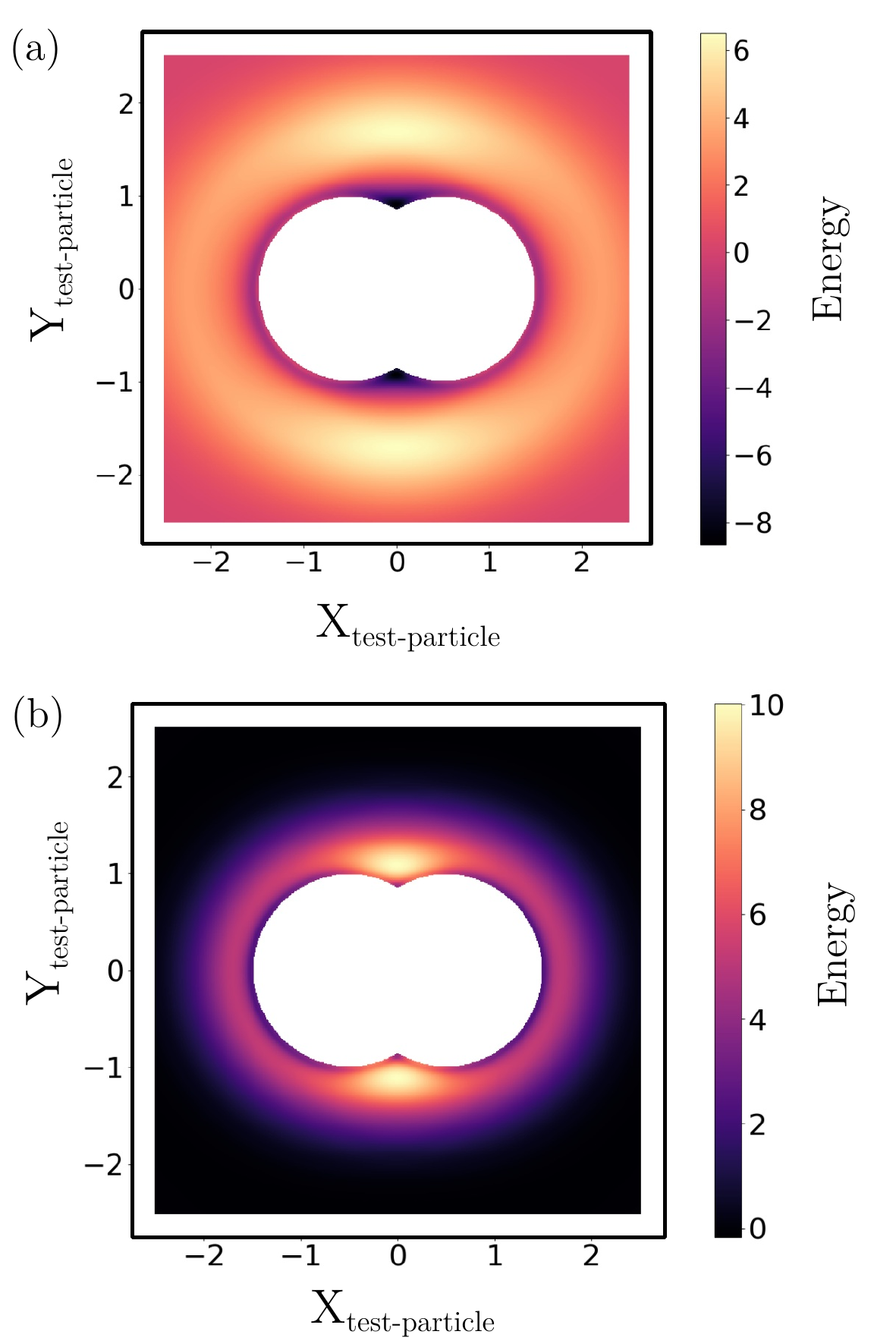}
        \caption{Two-dimensional potential energy landscape around a dimer as viewed by a test particle using the optimized potential for (a) $N_{\text{tgt}} = 4$ compact clusters at $\eta$ = 0.025 and (b) chainlike clusters at $\eta = 0.15$. The $X$ and $Y$ axes (in units of $\sigma$) denote $x$ and $y$ coordinates of the test particle while the color bar is in units of $k_{B}T$.}
        \label{fgr:heatmap}
    \end{figure}

    Self-assembly is the result of an interplay of energetic and entropic contributions that determine which types of structures minimize the free energy of a given system. Despite this inherent complexity, we can gain insight into the propensity for a given potential to form either stringlike or compact objects by considering the energy for a test particle to approach a dimer, where ``end-attachment'' to the dimer gives rise to a short string and ``middle-attachment'' yields a compact triangle. A slice of the potential energy landscape seen by the test particle relative to an ideal dimer is shown in Fig.~\ref{fgr:heatmap} as a heat map for two illustrative cases: (a) the potential optimized for compact tetramers at $\eta$ = 0.025 and (b) the potential optimized for strings at $\eta$ = 0.15. The corresponding heat maps for the remaining cluster and string cases are demonstrated in Figs. A4 and A5 in the Appendix. For the compact cluster-forming potential, the energy is lowest when the test particle bonds to both particles of the dimer simultaneously due to the deep attractive well at $r=\sigma$, promoting middle-attachment. Because the potential has been optimized to form compact tetramers, the repulsive corona surrounding the pair of particles in Fig.~\ref{fgr:heatmap}a penalizes the middle- or end-attachment of additional particles necessary to form larger compact clusters or strings. 
    
    By contrast, when the potential optimized for string formation is used, end-attachment to the dimer is more energetically favorable than middle-attachment, fostering chain growth. End-attachment is favored for this particular interaction because the energy of the attractive well at $r=\sigma$ is greater than at $r=2\sigma$, in part due to the relatively narrow repulsive barrier. Unlike for clusters, the chain-forming potential is net repulsive so that the lowest energy position for the test particle is to not attach to the dimer at all. However, in a bulk system, sufficiently high pressures induce particle association. Note that the preceding preference for end-attachment over middle-attachment will continue to be present as particles are added to the chain; that is, there is no energetic mechanism for controlling the length of the chain inherent to the potential itself. 
    
    While the above analysis is limited in that potentially important effects (e.g., the impact of surrounding aggregates on the energetics, the role of entropy, etc.) are omitted, this simplified model lends insights into how the lengthscales of the attractive well and the repulsive barrier might influence self-assembly. In particular, the position of the repulsive barrier controls the size of the compact clusters, in keeping with prior work.~\cite{RyanIC} Furthermore, the relatively narrow repulsive barrier in the string-forming potential promotes end attachment. 
    
\section{Universal Potential}
\label{sec:UP}

    \begin{figure}
        \includegraphics[width=3.37in,keepaspectratio]{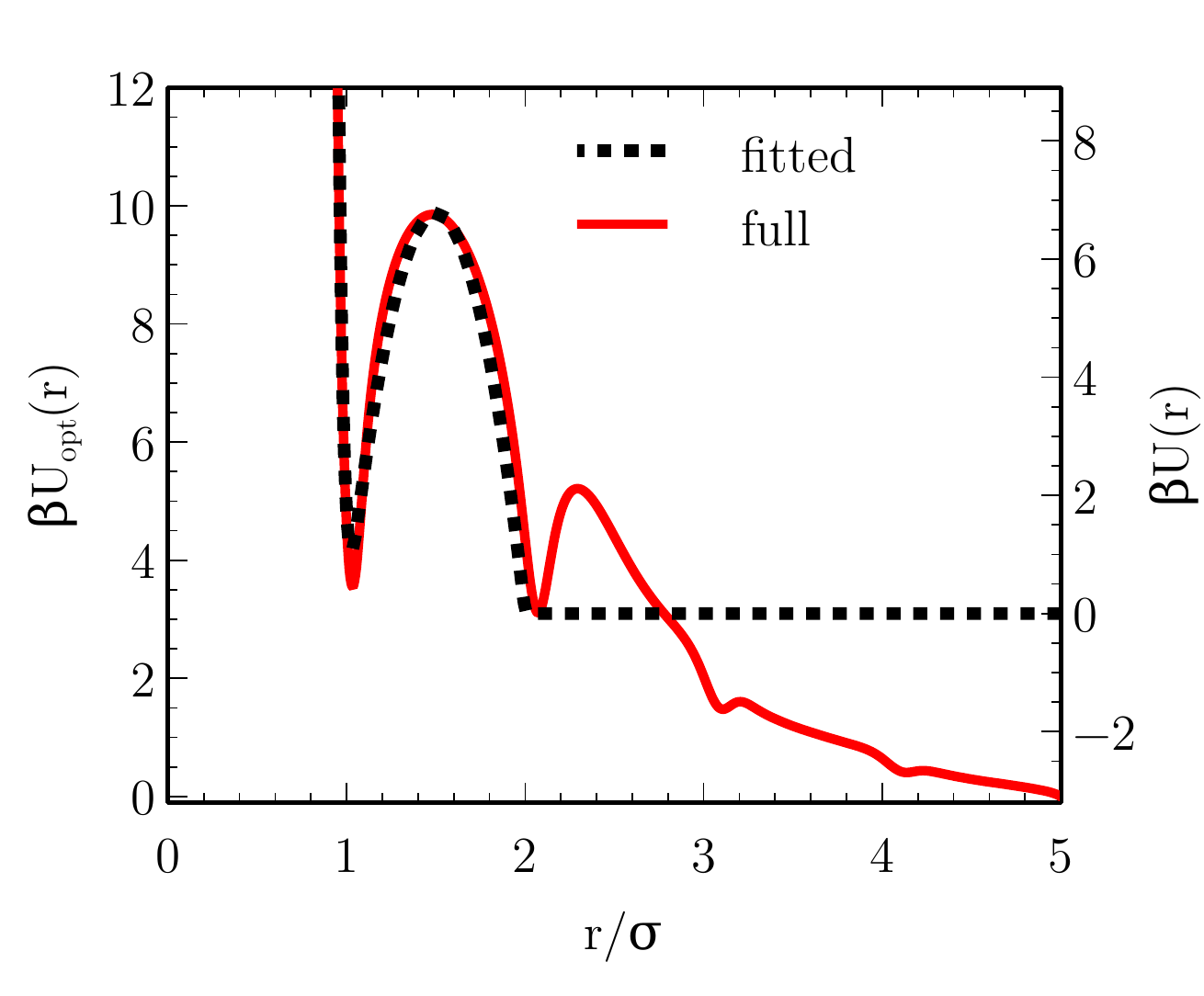}
        \caption{(solid) Optimized potential for strings at $\eta = 0.05$. (dotted) The fit of Eqn. 6-7 to the preceding potential after the potential has been truncated at $r = 2\sigma$ and shifted so that $U(r = 2\sigma)=0$. The optimal parameters are $\beta \varepsilon_{R} = 6.76$, $\delta_{A}=0.5$, and $\delta_{R}=0.5$}
        \label{fgr:cut_potn}
    \end{figure}

    Comparing the pair potentials optimized for compact clusters to those optimized for chains, the common features are an attractive well at short separations and an outer repulsive barrier with a shorter range than that characteristic of the Yukawa potential routinely used in SALR potentials. As noted in Sect.~\ref{subsec:IO_HeatMap}, the location of the repulsive barrier (which is controlled by the widths of the attractive well and the repulsive barrier) is one of the key parameters in promoting either end or middle attachment. 
    
    Motivated by these basic features shown in Figs.~\ref{fgr:RE_cluster} and~\ref{fgr:RE_string_potn}, we propose a simple and tunable pair potential form that favors various equilibrium structures, from disordered monomeric fluid to thin particle strings to compact particle clusters. Specifically, we consider a sum of a steep WCA repulsion at short distance followed by two half harmonic potentials that mimic the attractive well and the repulsive barrier.

    \begin{equation}
        \label{eqn:Potn}
            U(r) = \Phi_{\text{wca}}(r)+\Phi_{\text{hp}_{1}}(r)+\Phi_{\text{hp}_{2}}(r)
    \end{equation}
    
    The three sub-potentials are defined as
    
    \begin{equation}
        \label{eqn:WCA1}
            \begin{array}{ll}
            \Phi_{\text{wca}}(r) = \left\{
                             \begin{array}{ll}
                            4 {\varepsilon}_{\text{wca}}\bigg[ \bigg(\dfrac{\sigma}{r} \bigg)^{2\alpha} - \bigg( \dfrac{\sigma}{r} \bigg)^{\alpha} \bigg] + {\varepsilon}_{\text{wca}} & r \leq \sigma \\ [15pt]
            
                            0 & r > \sigma
                            \end{array}
                        \right.
            \end{array}
    \end{equation}            
    
    \begin{equation}
        \label{eqn:harmonic1}
            \begin{array}{ll}
            &\Phi_{\text{hp}_{1}}(r) = \left\{
                             \begin{array}{ll}
                             0 & r < \sigma \\ [15pt]
                        
                             \varepsilon_{\text{R}} \bigg[ 1 - \bigg(\dfrac{r - w_{1}}{\delta_{A}}\bigg)^2 \bigg] & \sigma \leq r \leq w_{1} \\ [15pt]
                        
                            0 & r > w_{1} \\ [15pt]
                            \end{array} 
                        \right. 
            \end{array}
    \end{equation}
                        
    \begin{equation}                    
        \label{eqn:harmonic2}
            \begin{array}{ll}
            \Phi_{\text{hp}_{2}}(r) = \left\{
                             \begin{array}{ll}
                             0 & r \leq w_{1} \\ [15pt]
                        
                             \varepsilon_{\text{R}} \bigg[ 1 - \bigg(\dfrac{r - w_{1}}{\delta_{R}}\bigg)^2 \bigg] & w_{1} \leq r \leq w_{2} \\ [15pt]
                        
                            0 & r \geq w_{2} \\ [15pt]
                            \end{array} 
                        \right. 
            \end{array}
    \end{equation}
    where $\beta \varepsilon_{wca} = 1.5$, $\alpha = 12$, $w_{1} = \sigma + \delta_{A}$, and $w_{2} = \sigma + \delta_{A} + \delta_{R}$. The remaining adjustable parameters ($\beta \varepsilon_{R}, \delta_{A}, \delta_{R}$) are determined by fitting the above form to the optimized potential that promotes self-assembly of strings at $\eta = 0.05$. This particular potential is used as the reference because strings are generally more challenging to self-assemble than compact clusters from an isotropic potential, and $\eta = 0.05$ is the packing fraction at which the designed potential resulted in strings with the smallest fractal dimension. To simplify the reference, the optimized potential is truncated beyond $r = 2\sigma$ and vertically shifted so that $U(r = 2\sigma)=0$. As shown in Fig.~\ref{fgr:cut_potn}, the resulting fit approximates the short-ranged ($r \leq 2\sigma$) reference potential well with $\beta \varepsilon_{R} = 6.76$, $\delta_{A}=0.5$, and $\delta_{R}=0.5$. To avoid discontinuities in the force profile, the above potential is weakly smoothed using a successive two-point averaging scheme where, beyond $r = \sigma$, $U(r_i)$ is twice replaced by an average of $U(r_{i-1})$ and $U(r_{i+1})$. 
    
    \subsection{Description of Morphologies}
    \label{subsec:morphologies}
    
\begin{figure}
        \includegraphics[width=3.37in,keepaspectratio]{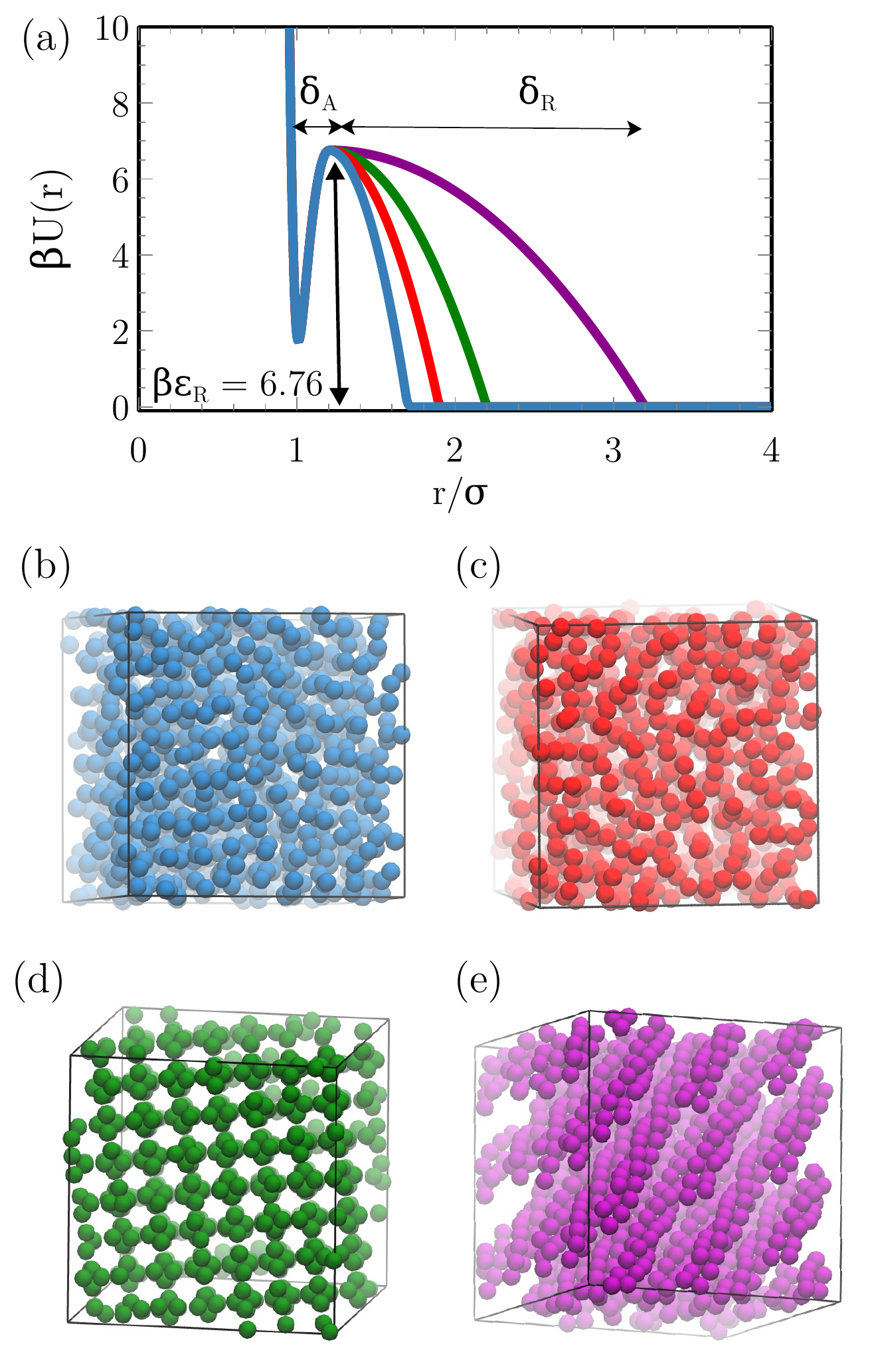}
        \caption{(a) Proposed model potential given by Eq.~\ref{eqn:Potn} for $\delta_{R}=0.5, 0.7, 1.0, 2.0$ at fixed $\delta_A=0.2$ and $\beta\varepsilon_{R} = 6.76$. Snapshots of assembled structures obtained via molecular dynamics simulations using the potentials shown in panel a, including (b) short strings ($\delta_{R} = 0.5$) (c) percolated chains ($\delta_{R} = 0.7$) (d) crystalline clusters ($\delta_{R} = 1.0$) and (e) Bernal spirals ($\delta_{R} = 2.0$).}
        \label{fgr:Snapshots}
    \end{figure}

\begin{figure}
        \includegraphics[width=3.37in,keepaspectratio]{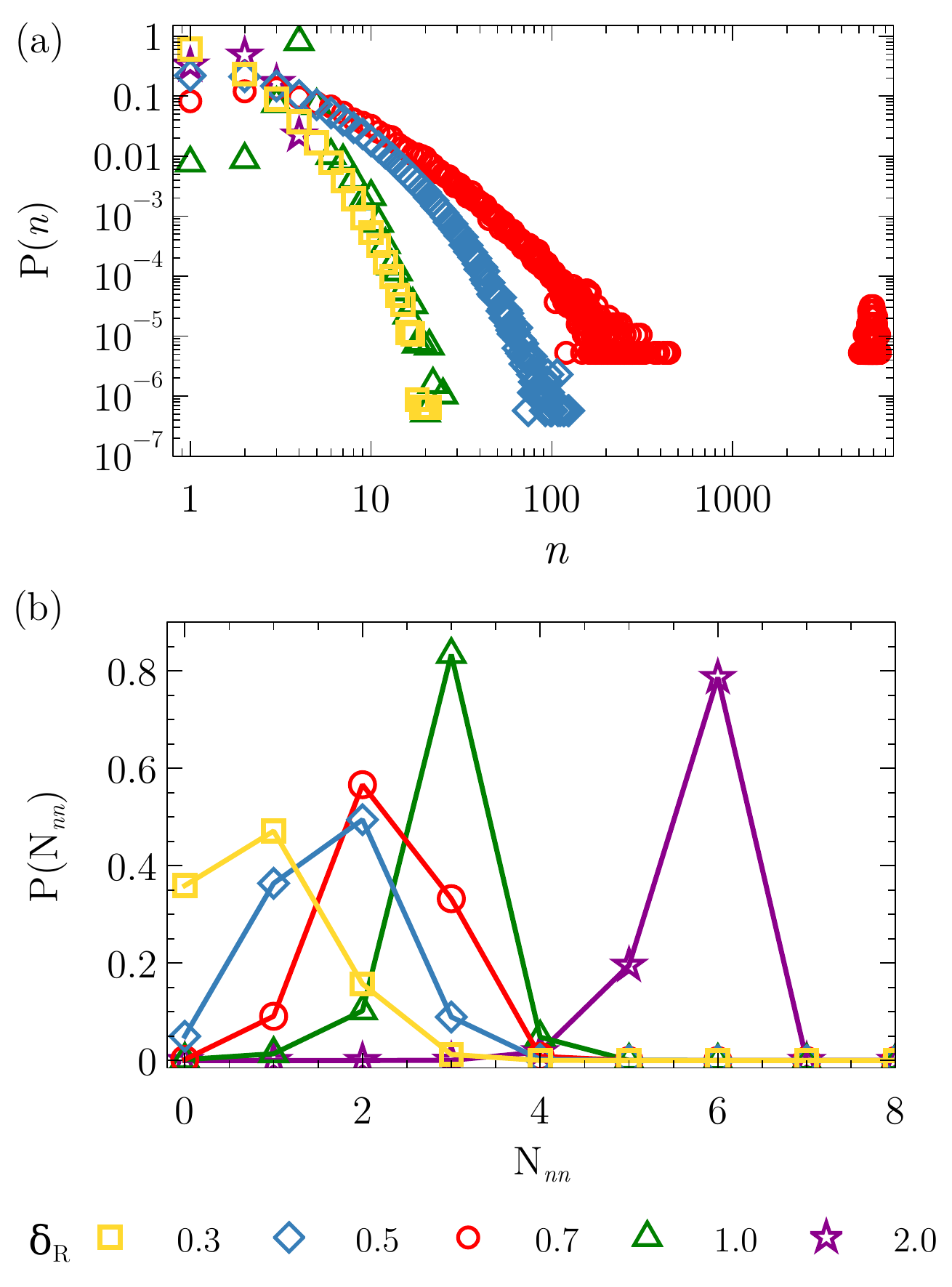}
        \caption{Using the model potential of Eqn.~\ref{eqn:Potn} at $\eta = 0.15$, $\delta_A = 0.2$, and different values of $\delta_R$, simulated (a) fraction of clusters containing $n$ particles, $P(n)$, and (b) distribution of the average number of nearest neighbours of each particle.}  
        \label{fgr:CSD}
    \end{figure}

    In this Section, we explore the effects of tuning the ranges of attraction ($\delta_{A}$) and repulsion ($\delta_{R}$) in the model pair potential introduced in Sect.~\ref{sec:UP}, while holding $\beta \varepsilon_{R}=6.76$ constant at a packing fraction of $\eta = 0.15$ (i.e., the lowest value of $\eta$ for which we observed the percolated string morphologies described below). Based on the discussion of Sect.~\ref{subsec:IO_HeatMap}, we expect that modifying $\delta_{A}$ and $\delta_{R}$ will bias the potential toward favoring assembly of either strings or compact clusters. Fig.~\ref{fgr:Snapshots}a shows four possible potentials where $\delta_{A} = 0.2$ and $\delta_{R} = 0.5, 0.7, 1.0, 2.0$. For the family of potentials where only $\delta_{R}$ is varied while $\delta_{A} = 0.2$ is constant, we observe four broad classes of structures in molecular dynamics simulations, shown in Fig.~\ref{fgr:Snapshots}b-e and with corresponding CSDs and nearest neighbour distributions shown in Figs.~\ref{fgr:CSD}a and b, respectively.
    
    \begin{enumerate}
    
        \item \textbf{Monomers (M)} A fluid of particles that remain in a dispersed state results when $\delta_{A} = 0.2$ and $\delta_{R} = 0.3$. Quantitatively, a state is defined to be monomeric if the fraction of monomers (cluster size $n = 1$) in the CSD exceeds 50\%. The CSD for this state point (Fig.~\ref{fgr:CSD}a) shows 22\% of the aggregates are dimers, while 65\% are monomers. Correspondingly, the nearest neighbour histogram (Fig.~\ref{fgr:CSD}b) shows that a significant majority of the particles have zero or one nearest neighbor.  
    
        \item \textbf{Strings (S and SP)} Single-stranded stringy particle assemblies are obtained at values for $\{\delta_{A}$, $\delta_{R}\}$ of $\{0.2, 0.5\}$ and $\{0.2, 0.7\}$. We distinguish between shorter stringlike objects (S) and percolated networks of strings (SP) on the basis of the percolation analysis described in Sect.~\ref{subsec:StructuralAnalysis}. Snapshots of short strings ($\delta_{R}=0.5$) and percolated chains ($\delta_{R}=0.7$) are depicted in Figs.~\ref{fgr:Snapshots}b and c respectively. Stringlike structures, regardless of whether percolated, have predominantly two bonds per particle. The primary difference upon transitioning from shorter strings to a percolated porous network is the growth in the number of junctions, imparting greater connectivity to the branches. Thus, the corresponding nearest neighbour distribution for the percolated network shows a higher value for $N_{nn} = 3$ as compared to that of strings; see Fig.~\ref{fgr:CSD}b. For short strings, 
        we confirm that the aggregates are elongated on the basis of visual inspection and the $d_f$. As observed for the pair potential optimized for strings at $\eta=0.15$ with $\{\delta_{A}$, $\delta_{R}\}$ values of $\{0.2, 0.5\}$, we observed two distinct regimes of $d_f$ as a function of aggregate size $n$. For $n \leq 11$, $d_f=1.35$, and for larger aggregates $d_f=1.9$. 
        More generally, we identify structures as stringlike on the basis of both a peak at $P(N_{nn})=2$ and a $d_f$ in the range of $1-1.5$ for small ($n \lesssim 10$) objects and a $d_f$ around $1.9-2.0$ for the longer chains. 
        
        Compared to the results presented in Sect.~\ref{subsec:IO_Strings}, there are markedly fewer compact aggregates in coexistence with the strings: 1\% of aggregates for the short strings and 0.5\% of aggregates for the percolated strings are compact. However, in keeping with the inversely designed potentials, there is no apparent size-specificity for the strings. The CSDs in Fig.~\ref{fgr:CSD}a show a monotonic trend of decreasing probability of larger aggregates for the unpercolated structures. To the best of our knowledge, this is the first demonstration of the self-assembly and stabilization of spherical particles into a three dimensional open, porous network of \textit{single-stranded} chains via an isotropic potential at nonzero temperature.

        \item \textbf{Clusters (C)} Well-defined compact clusters are observed (Fig.~\ref{fgr:Snapshots}d) using the proposed potential at $\{\delta_{A}$, $\delta_{R}\}=\{0.2, 1.0\}$. Clusters are identified by a sharp maximum in the CSD at a cluster size $n > 1$ and by a fractal dimension of $\lesssim$ 3 owing to their compact and isotropic shape. At this state point, the clusters are tetramers; see the prominent peaks at $P(n)=4$ and $P(N_{nn})=3$ in Fig.~\ref{fgr:CSD}a,b. The corresponding $d_f$ is 2.5. Instead of a fluid of clusters as studied in Sect.~\ref{subsec:IO_Clusters}, the self-assembled clusters crystallize under these conditions onto a lattice.
        
        \item \textbf{Cylindrical Spirals (CS)} Elongated, cylindrical spirals of colloids are observed at $\{\delta_{A}$, $\delta_{R}\}=\{0.2, 2.0\}$. These are multi-stranded, percolated networks of anisotropic aggregates. A special case of these structures with three helical chains is commonly referred to as Bernal spirals ~\cite{Bartlett, Wales, OneD, Bolhuis} which have six nearest neighbours, and a representative snapshot is shown in Fig.~\ref{fgr:Snapshots}e. Accordingly, Bernal spirals are identified as percolated structures with a peak in $P(N_{nn})=6$.
    \end{enumerate}

    \begin{figure}
        \includegraphics[width=3.37in,keepaspectratio]{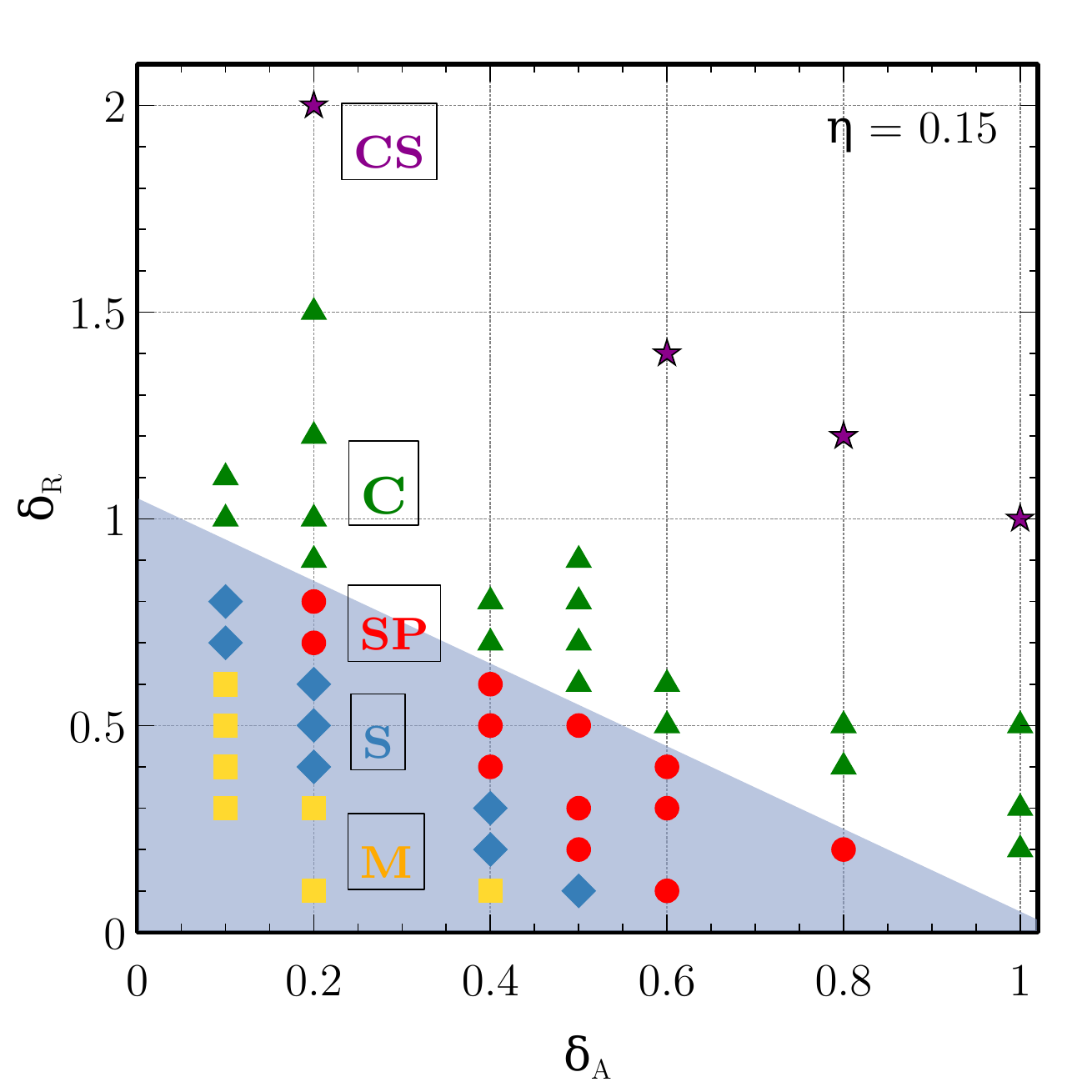}
        \caption{Morphological phase diagram for the model potential of Eqn.~\ref{eqn:Potn} as a function of $\delta_{A}$ and $\delta_{R}$ at $\eta = 0.15$ and $\varepsilon_{R} = 6.76$. The structures observed include monomers M (square), short strings S (diamonds), percolated strings SP (circles), crystalline clusters (triangles), and Bernal Spirals CS (stars) discussed in the text. The shaded region depicts phase space where end-to-end joining of monomers--needed for "thin" strings--is favored over more compact aggregate packings.}  
        \label{fgr:PhaseDiagram}
    \end{figure}   
   
    \subsection{Phase Diagram}
    \label{subsec:phase_diagram}
   
   In Fig.~\ref{fgr:PhaseDiagram}, we further explore how the morphologies identified above emerge in this model as a function of $\delta_R$ and $\delta_A$. For the conditions studied, note that the quantity $\delta_{A} + \delta_{R}$ appears to be the primary determinant of which self-assembled structures are observed. When $\delta_{A} + \delta_{R} \lesssim 0.5$, the particles form a fluid of well-dispersed monomers (M). For progressively larger $\delta_{A} + \delta_{R}$, short single-stranded strings (S) of particles form followed by interconnected percolating networks of strings (SP). For sufficiently low attractive ranges ($\delta_A \leq 0.2$), the physical bonds are labile and percolated strings are fluidic, continually breaking and reforming flexible uniaxial structures. However, for $\delta_A > \delta_R$, the increased attraction impedes dissociation of the particles especially at the junctions. As a result, though the chains still fluctuate in local order, larger length-scale motions are suppressed. 

    When $\delta_R + \delta_A \gtrsim 1$, the space-spanning stringy particle network morphs into a crystalline arrangement of compact particle clusters (C). Consistent with Fig.~\ref{fgr:RE_cluster}a, the clusters grow with increasing $\delta_A$. For $\delta_A \lesssim 0.6$, the clusters are tetramers, but for larger values of $\delta_A$, aggregation numbers of $5-7$ are observed. The corresponding $d_f$ values range from $2.5-2.6$.  Increasing $\delta_R + \delta_A$ further compels the spherical clusters to coalesce, eventually resulting in kinetically arrested percolated networks of cylindrical structures (CS), of which the Bernal Spiral is a special case indicated in Fig.~\ref{fgr:PhaseDiagram}. 
        
    As detailed above, the interplay between the length scales of the attraction and repulsion in this model pair potential results in a rich variety of self-assembled structures. The analysis in Sect.~\ref{subsec:IO_HeatMap} suggested that comparing the energy for a test particle to bond to either end of a dimer versus the mid-point is a helpful, though simplistic, predictor of whether the clusters formed by a given potential will be stringy or compact, respectively. Here, we also find that this analysis helps to understand why $\delta_R + \delta_A$ is an important parameter in determining the observed morphologies. Specifically, we compare the energetics of end-attachment ($U_{\text{end}} \equiv U(\text{r}=\sigma) + U(\text{r}=2\sigma)$) and middle-attachment ($U_{\text{mid}} \equiv U(\text{r}=\sigma) + U(\text{r}=\sigma)$) of a test particle to an isolated dimer. Regions of phase space where end-attachment is preferred ($U_{\text{end}} \leq U_{\text{mid}}$) are shaded in Fig.~\ref{fgr:PhaseDiagram}. Note that when $\delta_{A}+\delta_{R} \lesssim 1.0$, end-attachment to a dimer is preferred; when $\delta_{A}+\delta_{R}$ is larger, middle-attachment is favored. Interestingly, $\delta_{A}+\delta_{R} \approx 1.0$ also approximately corresponds to the crossover between percolated string networks and compact clusters observed in the simulations. Thus, though based on a simplistic energetic analysis, we can gain insights into how the length scales of the short-range attractions and longer-range repulsions in a pair potential can favor the formation of stringlike versus compact self-assembled structures.

\section{Conclusions}
We used an inverse design strategy based on RE optimization to determine and study isotropic potentials capable of driving one-component systems of particles to self-assemble into compact versus linear stringlike clusters. Simulations using particles interacting via the optimized potentials demonstrated spontaneous formation of the targeted morphologies, though successful design of specific aggregation numbers was only achievable for compact clusters, a limitation that might be expected for isotropic potentials. The simplicity of the optimized potentials for these structures is remarkable given that prior computational efforts to arrive at stringlike clusters in three dimensions have employed directional bonding or anisotropy of the colloidal building blocks to control assembly. 

Motivated by the RE optimized potentials, a universal potential with a simple functional form is proposed that is capable of assembling a rich variety of complex architectures:  monomeric fluid, fluid of short chain-like structures, percolated networks of strings, crystalline assemblies of compact clusters, and percolated thick cylindrical structures including Bernal spirals. The proposed model potential is a combination of short-range attraction at contact, which can be realized by polymer-mediated depletion in chemistry-matched systems (for smaller values of $\delta_{A}$) and a medium-ranged repulsive barrier which approximately mimics that of suspensions of non-charged brush-grafted nanoparticles.~\cite{Denton} Polymer depletants that are responsive to external stimuli (e.g., pH,~\cite{pH_polymer} temperature,~\cite{thermo_polymer} light~\cite{photo_polymer} and other fields~\cite{field_polymer}) represent another interesting avenue to tune $\delta_{A}$ to switch between different morphologies. More generally, these results provide qualitative insights into the rich morphological phase diagrams that can potentially be realized in colloidal systems with (approximately) isotropic interactions with competitive repulsive and attractive components.

\label{sec:conclusions}

\section*{Acknowledgments}
The authors thank Sanket Kandulkar and Michael P. Howard for valuable discussions and feedback. This research was primarily supported by the National Science Foundation through the Center for Dynamics and Control of Materials: an NSF MRSEC under Cooperative Agreement No. DMR-1720595 as well as the Welch Foundation (F-1696). We acknowledge the Texas Advanced Computing Center (TACC) at The University of Texas at Austin for providing HPC resources.

%


\setcounter{figure}{0}
\setcounter{equation}{0}
\renewcommand\thefigure{A\arabic{figure}}
\renewcommand{\thesection}{\thepart .\arabic{section}}
\renewcommand\theequation{A\arabic{equation}}
\renewcommand{\thesubsection}{\arabic{subsection}}
\renewcommand{\thesubsubsection}{\alph{subsubsection}}

\section*{Appendix}

    \begin{figure*}[!h]
    \centering
        \includegraphics[width=3.37in,keepaspectratio]{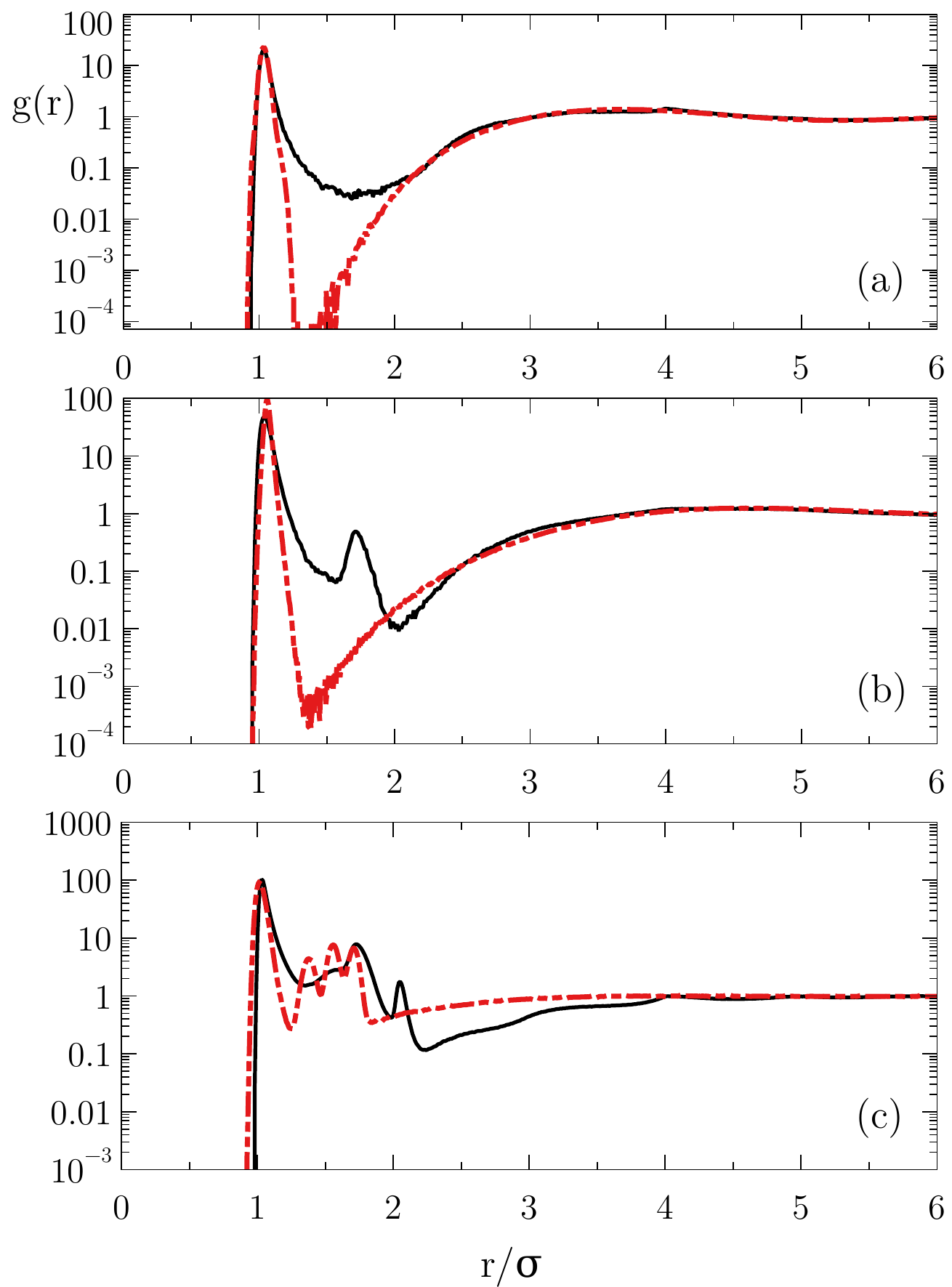}
        \caption{Comparison of the target (dot-dashed) and the optimized (solid) radial distribution functions $g{\text{(r)}}$ for the compact clusters of size $N_{\text{tgt}} =$ (a) 2, (b) 4, and (c) 8.}
        \label{fgr:rdf_cluster}
    \end{figure*}

    \begin{figure*}[!h]
        \includegraphics[width=3.37in,keepaspectratio]{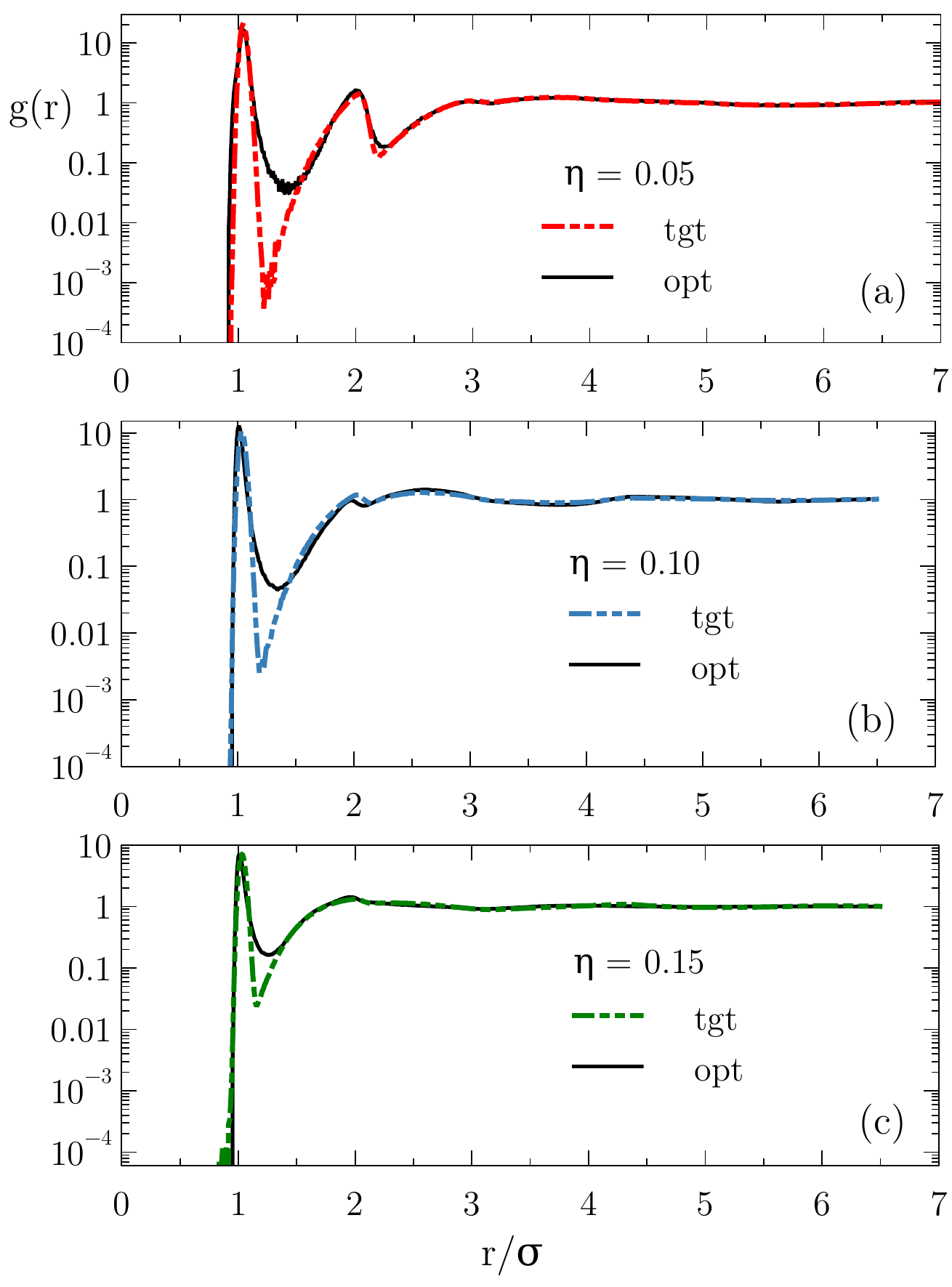}
        \caption{Comparison of the target and the optimized radial distribution functions $g{\text{(r)}}$ for the strings of volume fraction $\eta = $ 0.05, 0.10, and 0.15.}
        \label{fgr:rdf_string}
    \end{figure*}
    
    \begin{figure*}[!h]
        \includegraphics[width=3.37in,keepaspectratio]{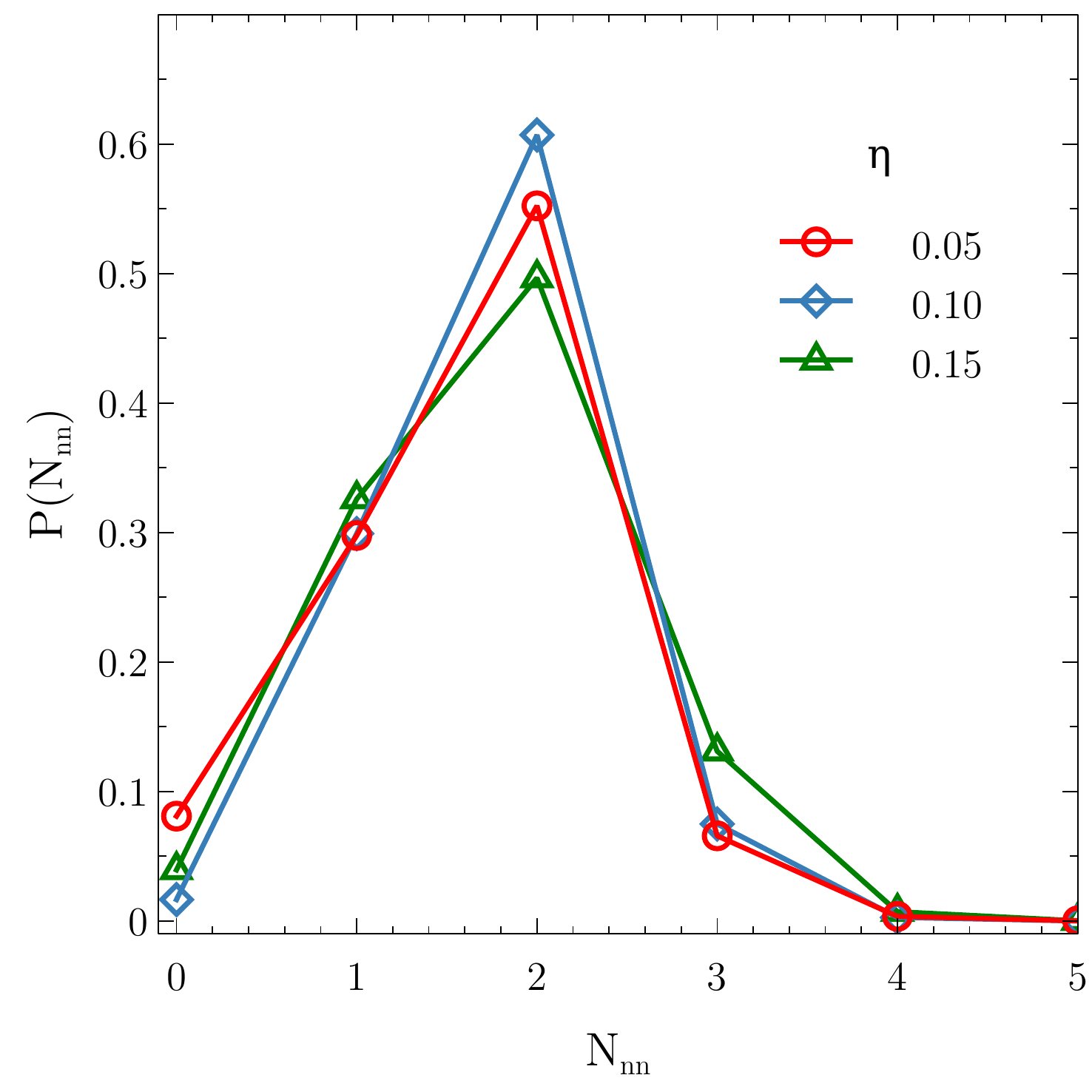}
        \caption{Distribution of the number of nearest neighbours at three different volume fractions with the compact clusters of size $3 \leq n \leq 5$ removed.}
        \label{fgr:nnhist_REstring_compactREMOVED}
    \end{figure*}
    
     \begin{figure*}[!h]
        \includegraphics[width=3.37in,keepaspectratio]{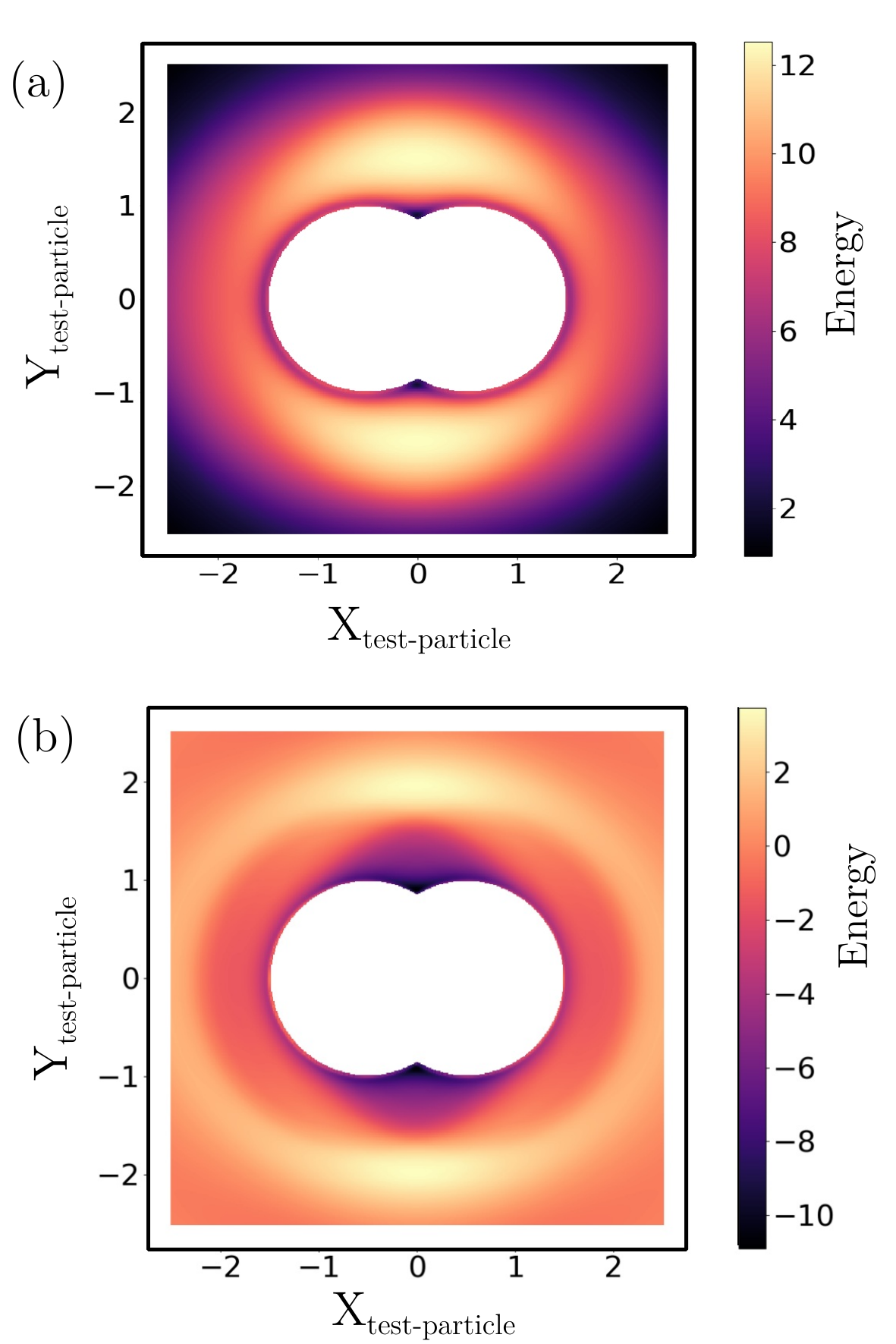}
        \caption{Two-dimensional potential energy landscape around a dimer as viewed by a test particle using the optimized potential for compact clusters of size (a) $N_{\text{tgt}}$ = 2 and (b) $N_{\text{tgt}}$ = 8. The x and y axes (in units of $\sigma$) denote coordinates of the test particle while the color bar is in units of $k_{B}T$.}
        \label{fgr:SI_cluster_heatmap}
    \end{figure*}

    \begin{figure*}[!h]
        \includegraphics[width=3.37in,keepaspectratio]{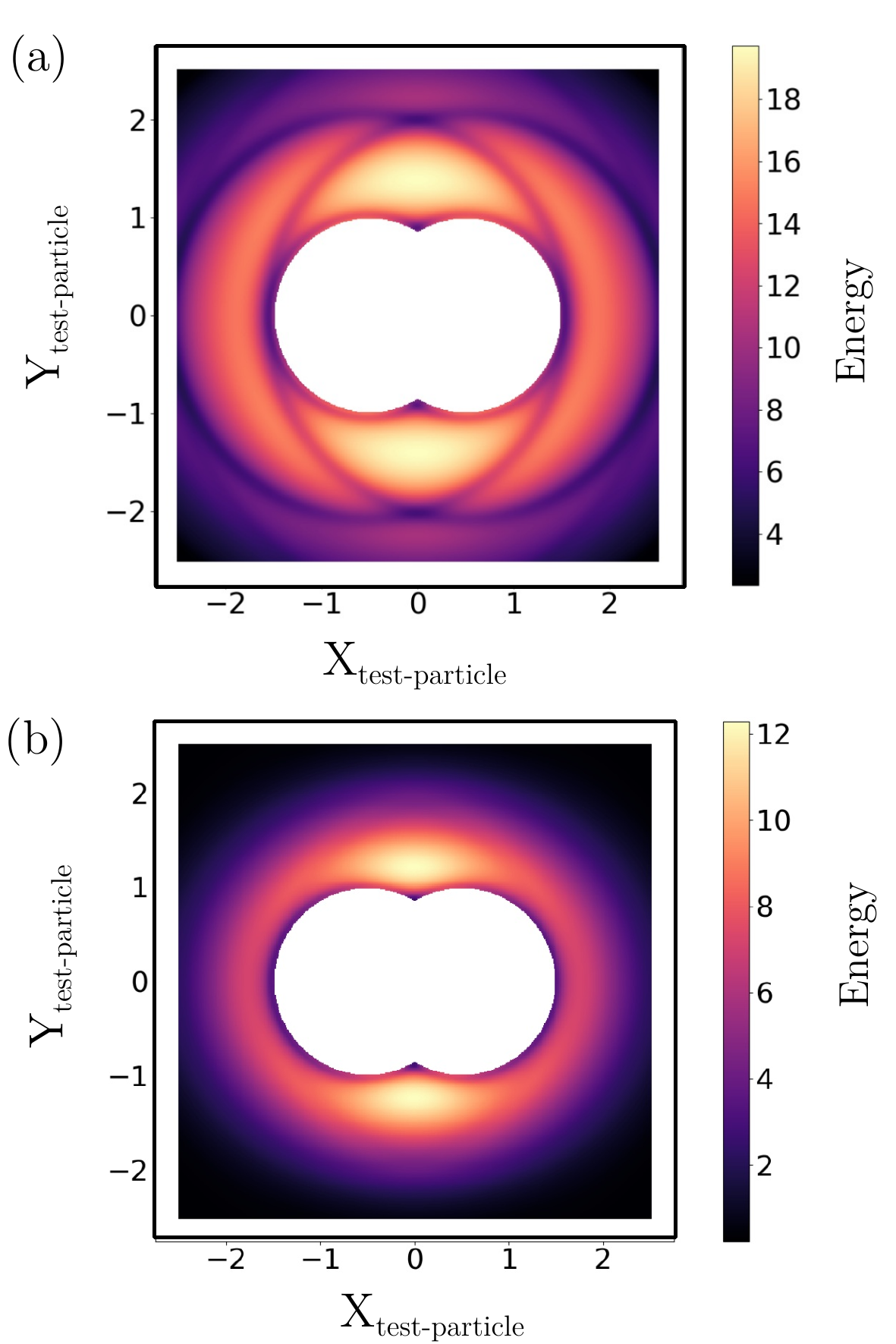}
        \caption{Two-dimensional potential energy landscape around a dimer as viewed by a test particle using the optimized potential for particle strings at (a) $\eta$ = 0.05 and (b) $\eta = 0.10$. The x and y axes (in units of $\sigma$) denote coordinates of the test particle while the color bar is in units of $k_{B}T$.}
        \label{fgr:SI_string_heatmap}
    \end{figure*}


\end{document}